\documentclass[AMA,STIX1COL]{WileyNJD-v2}

\usepackage{graphicx}
\usepackage{subfig}
\usepackage{hyperref}

\articletype{Research Article}%

\received{XXXX}
\revised{XXXX}
\accepted{XXXX}

\begin{document}

\title{BigDataSDNSim: A Simulator for Analyzing Big Data Applications in Software-Defined Cloud Data Centers}

\author[1]{Khaled Alwasel}
\author[2]{Rodrigo N. Calheiros}
\author[3]{Saurabh Garg}
\author[4]{Rajkumar Buyya}
\author[1]{Rajiv Ranjan}

\authormark{Alwasel \textsc{et al}}

\address[1]{\orgdiv{School of Computing}, \orgname{Newcastle University}, \orgaddress{\state{Newcastle upon Tyne}, \country{UK}}}

\address[2]{\orgdiv{School of Computing, Engineering and Mathematics}, \orgname{Western Sydney University}, \orgaddress{\state{Penrith}, \country{Australia}}}

\address[3]{\orgdiv{School of Computing and Information Systems}, \orgname{University of Tasmania}, \orgaddress{\state{Hobart}, \country{Australia}}}

\address[4]{\orgdiv{School of Computing and Information Systems}, \orgname{The University of Melbourne}, \orgaddress{\state{Melbourne}, \country{Australia}}}

\corres{Khaled Alwasel, School of Computing, Newcastle University, UK.\newline\email{kalwasel@gmail.com}}

\abstract[Summary]{Emerging paradigms of big data and Software-Defined Networking (SDN) in cloud data centers have gained significant attention from industry and academia. The integration and coordination of big data and SDN are required to improve the application and network performance of big data applications. While empirical evaluation and analysis of big data and SDN can be one way of observing proposed solutions, it is often impractical or difficult to apply for several reasons, such as expensive undertakings, time consuming, and complexity; in addition, it is beyond the reach of most individuals. Thus, simulation tools are preferable options for performing cost-effective, scalable experimentation in a controllable, repeatable, and configurable manner. To fill this gap, we present a new, self-contained simulation tool named BigDataSDNSim that enables the modeling and simulating of big data management systems (YARN), big data programming models (MapReduce), and SDN-enabled networks within cloud computing environments. To demonstrate the efficacy, effectiveness, and features of BigDataSDNSim, a use-case that compares SDN-enabled networks with legacy networks in terms of the performance and energy consumption is presented.}


\jnlcitation{\cname{%
\author{Alwasel K}, 
\author{Calheiros RN}, 
\author{Garg S}, 
\author{Buyya R}, and 
\author{Ranjan R.}}
\ctitle{BigDataSDNSim: A Simulator for Analyzing Big Data Applications in Software-Defined Cloud Data Centers}, \cjournal{Softw Pract Exper}, \cvol{2018;00:1--6}.}

\maketitle

\section{Introduction}\label{introduction}
\begin{figure}[t] 
	\centering
	{\includegraphics[]{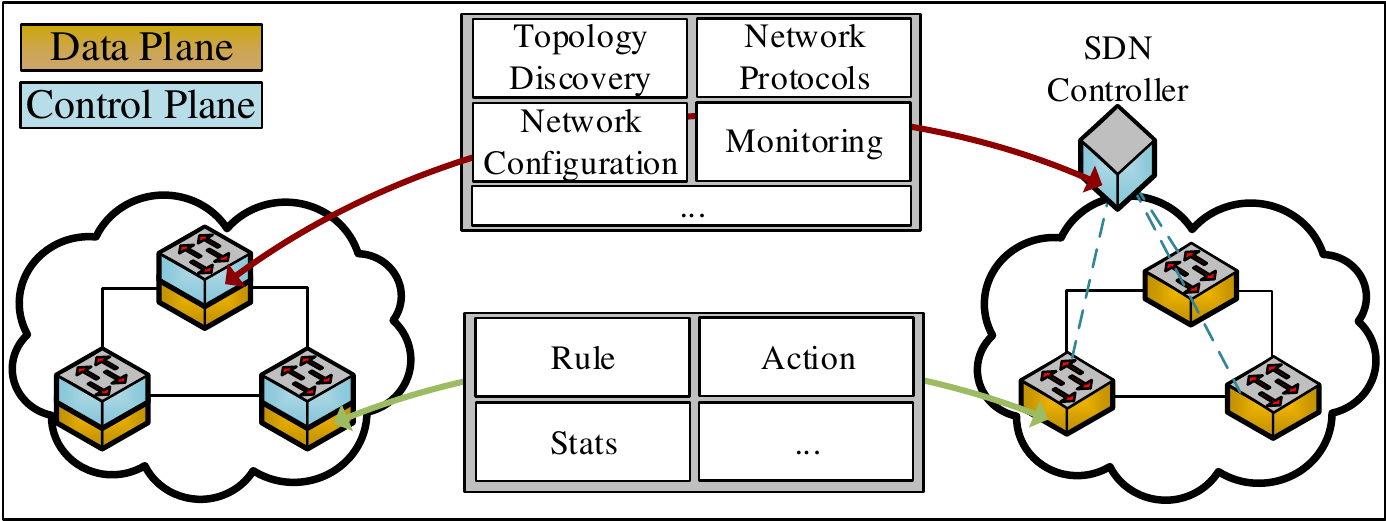}}
	\caption{Key difference between traditional network and SDN-enabled network.}
	\label{fig_SDN_VS_Legacy}%
\end{figure}

The Software-Defined Networking (SDN)\cite{kreutz2015software}  paradigm aims to enable dynamic configuration of networks and overcome the limitations of current network infrastructure. It decouples the control layer (configuration and routing policies of network devices) from the data layer (traffic forwarding) in contrast to traditional networks (see Figure \ref{fig_SDN_VS_Legacy}). Such abstraction makes SDN more robust, simplified and flexible to changes, traffic load, and faults. The control layer enforces the network management and routing policies in the data layer using well-defined application programming interfaces (APIs), where OpenFlow\cite{mckeown2008openflow} is the prominent protocol that provides such APIs. Because of these attractive features, SDN has recently gained significant attention from industry and academia. As a result, several network vendors (e.g. HP, Pica8, Netgear) have adopted the OpenFlow protocol and started to support its APIs in their network devices.  
 
Recently, the cloud community has started embracing the SDN paradigm to meet the demands of network-hungry big data applications. To effectively and efficiently process large volumes of big data, many big data frameworks (e.g., Apache Hadoop , Apache Storm , etc.) have been developed. Each framework deals with different aspects of big data requirements. For example, a MapReduce programming model analyses data in simultaneous and stateless fashions, while a stream processing model handles continuous big data streams in consecutive and stateful manners [2]. By using the SDN, big data applications can enforce their QoS requirements to the network layer on the fly along with accelerating data processing.

Despite the developments in both SDN and big data, several challenges still need to be tackled to leverage and evaluate the benefits of SDN for supporting the network capabilities for big data applications. The evaluation and analysis of various proposed solutions using live production environments can be impractical or difficult to apply for several reasons, such as expensive undertakings, time consuming, complexity, and instability. In order to address these challenges, simulation and emulation environments are required for performing cost-effective and scalable experimentation of proposed solutions. To fill this gap, we present a new, self-contained simulation tool named BigDataSDNSim that enables the modeling and simulating of big data management systems (YARN), big data programming models (MapReduce), and SDN-enabled networks within cloud computing environments.


The representation of real-world systems in simulation-based environments allows developers/researchers to observe, analyze, and determine the strengths and limitations of their hypothesis, design choices and proposed solutions. Simulation tools have the ability to execute large-scale infrastructure experiments easily, repeatedly, and in a controllable and configurable environment. In the last few years, there have been several efforts to design simulation and modeling environments. For example, Mininet\cite{lantz2010network} enables the emulation of SDN-based network where it implements customized routing protocols along with OpenFlow interfaces for instructing switches. Similarly, EstiNet\cite{wang2013estinet} allows both simulation and emulation of SDN. Both of these are good for solving network level challenges in SDN, but not at the application level where network traffic shape can be customized according to applications. 

There are also some efforts to integrate SDN with cloud simulators in order to monitor and optimize resources (e.g., CloudSimSDN\cite{son2015cloudsimsdn}). However, the application perspective of big data and the network perspective of SDN scheduling and management capabilities are still missing. In this context and to fill this gap, we present a new simulation tool named BigDataSDNSim that enables the modeling and simulation of big data management systems (e.g. YARN\cite{vavilapalli2013apache}) and big data programming models (e.g. MapReduce\cite{dean2008mapreduce}) in SDN-enabled cloud computing environments.

BigDataSDNSim is designed and implemented on top of CloudSim,\cite{calheiros2011cloudsim} in addition to the use of some functions of CloudSimSDN and IoTSim\cite{zeng2017iotsim} tools. It provides the required infrastructures and capabilities for evaluating and analyzing big data applications that utilize MapReduce programming models in SDN-powered cloud data centers. BigDataSDNSim is capable of quantifying the performance impacts of MapReduce applications by means of network and application design choices and scheduling policies coupled with other configuration factors. The contributions of this paper lie in the developing of a novel simulator that support following unique resource, network, and application programming model:
of (1) a big data management system (BDMS) that coordinates and schedules resources among competing big data applications co-hosted on shared cloud resources; (2) MapReduce big data programming model; (3) Software Defined Networking model (SDN); (4) SDN coordination and interaction with other cloud hosted system and application components (storage, application masters, mappers, etc.); (5) several SDN and application policies for multilevel optimization, reducing the complexity of deployment, and accelerating the process of evaluation and testing; and (6) dynamic routing mechanisms where any type of SDN topology can be seamlessly and efficiently simulated.

The rest of this paper is organized as follows: Section \ref{related_work} presents related work and reflects the importance and unique capabilities of our simulator. Section \ref{sec_design_implementation} elaborates the design and implementation of BigDataSDNSim framework in detail and demonstrates the features of built-in module policies. Section \ref{sec_evaluation} shows the simulation results and the efficiency, effectiveness, and practicality of our proposed simulator. Section \ref{sec_conclusion} concludes the paper and  highlights some future work. 

\begin{table}[b]
\centering
\caption{a comparison with existing simulators\\
* with the help of the INET framework\cite{INET}
; ** with the help of SDN controllers from other projects (e.g., Floodlight\cite{Floodlight})} \label{foo}

\includegraphics[clip,trim=3cm 38cm 3cm 3.1cm , width=17.85cm]{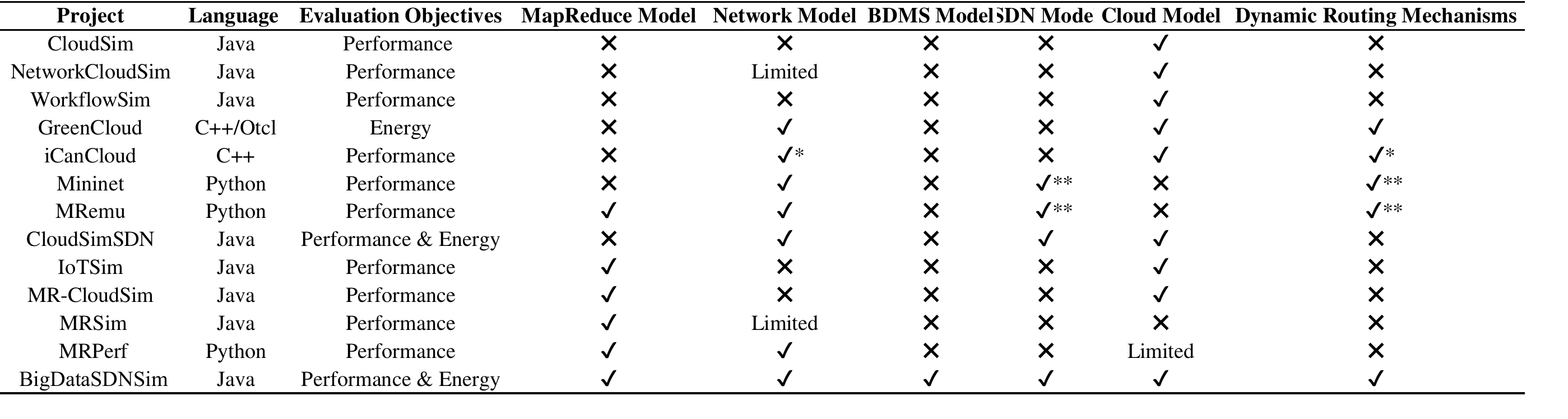}
\label{table_Comparison_related_work}%

\end{table}

\section{Related work}\label{related_work}
In recent years, simulation has become a powerful approach allowing researchers/developers to conduct in-depth evaluation and analysis of new algorithms and techniques (e.g., task scheduling strategies, energy conservation techniques, and VM allocation). To the best of our knowledge, state-of-the-art simulators are not capable of providing big data, SDN, and cloud infrastructures/models under one umbrella. Nevertheless, there are some existing simulation tools that can be related to our proposed simulator. Table \ref{table_Comparison_related_work} compares the characteristics and features of the existing simulators.

There are several simulation tools capable of simulating and modeling the behaviors and characteristics of cloud and legacy networks. For example, CloudSim \cite{calheiros2011cloudsim} is one of the most popular cloud-based tools that allows the modeling of physical and virtual cloud infrastructures using an event-driven architecture. It is capable of simulating and evaluating the performance of cloud infrastructures as well as deploying various provisioning and allocation policies (e.g. VM placement, CPU task scheduling). CloudSim forms the base upon which several simulation tools were developed to fill the gap of networks and applications aspects (e.g. NetworkCloudSim\cite{garg2011networkcloudsim}, WorkflowSim\cite{chen2012workflowsim}). GreenCloud\cite{kliazovich2012greencloud}, iCanCloud\cite{nunez2012icancloud}, and NetworkCloudSim are other cloud-based tools focusing on the perspective of legacy networks in terms of characteristics and communications in cloud data centers. However, these simulators lack (i) modeling and implementation of SDN-enabled network infrastructures (SDN controllers, OpenFlow mechanisms, dynamic routing algorithms, etc.); and (ii) constructs representing big data management systems (e.g. YARN) as well as big data programming models (e.g. MapReduce) where they impose strict behaviors and characteristics in terms of components' interactions, distributed-task dependencies, and data workflows.  

Mininet\cite{lantz2010network} is an SDN-based emulation tool that runs on a single device configured with a Linux-based system. It emulates different types of network topologies together with hundreds of virtual hosts and diverse UDP/TCP traffic patterns. The external SDN controller(s) communicates and enforces network policies on Mininet via its unique IP address and OpenFlow APIs. The main limitation of Mininet is that it does not support application-level infrastructures. To allow Mininet to mimic the behaviors of MapReduce applications, MRemu\cite{neves2015mremu} was introduced where it is capable of using realistic MapReduce workloads/traces within Mininet environments. It operates based on latency periods extracted from MapReduce job traces (duration of tasks, waiting times, etc.). Similarly, EstiNet\cite{wang2013estinet} is another SDN-based simulation and emulation tool allowing each simulated host to run a real Linux-based system coupled with several types of real applications. It is capable of simulating hundreds of OpenFlow switches and generating real TCP/IP traffic. However, none of these tools is capable of capturing details of big data management systems (e.g. YARN) in terms of system elements (application masters, node managers, etc.). In addition, they lack support for modeling of cloud-specific and MapReduce features, such as job placement policies, VM task scheduling policies, and VM placement policies. 

CloudSimSDN\cite{son2015cloudsimsdn} is an SDN-based cloud simulator that is modeled and implemented on top of CloudSim\cite{calheiros2011cloudsim}. It enables the simulation of SDN-network behaviors (SDN controllers, switches, links, packets, etc.) and measures the network performance based on allocated routing bandwidth (from a source VM to a destination VM) and latency of every link along the route. The focus of CloudSimSDN is towards reducing host-network energy consumption via joint power management policies. However, it fails to provide customizable network infrastructures; it has an embedded hard-coded implementation of three-tier fat-tree topology, which does not allow any other types of topologies to be automatically implemented (e.g., mesh and other tree topologies). Moreover, it also has limitations in (i) dynamic routing mechanisms for finding routes among sources and destinations on the fly; (ii) capabilities for deploying traffic policies for VMs/applications/jobs, such as prioritizing network access and bandwidth allocation; and (iii) an infrastructure of big data management systems and programming models. 

IoTSim \cite{zeng2017iotsim} is an extension of CloudSim that mimics the characteristics of the MapReduce programming model. It allows the simulation and modeling of Hadoop framework (job tracker, task trackers, mappers, and reducers) and provides a management mechanism to configure IoT applications. Similarly, MR-CloudSim\cite{jung2012mr}, MRSim\citep{hammoud2010mrsim}, and MRPerf\cite{wang2009using} are other tools that enable the simulation of MapReduce-based applications with different focuses and features. However, some noticeable deficiencies of these tools include (i) lack of generic approach of big data programming models and engines (e.g., YARN) where different types of programming models can be managed and executed simultaneously; (ii) lack of SDN/network mechanisms to optimize the network performance of diverse big data applications; and (iii) lack of big data management and execution policies to assure the efficiency and effectiveness of new techniques and approaches, such as job selection, job mapping, and map/reduce task scheduling.

\section{Architecture and Design}\label{sec_design_implementation}
In this section, we discuss the fundamental concepts, functionalities, and elements of BigDataSDNSim. The objective of our simulator is to offer a holistic integration of big data management system and big data programming models that are compatible with SDN network functions in cloud infrastructures. The logic of BigDataSDNSim is based on YARN (big data management system) and the MapReduce programming model, in  addition to the conceptually proposed architecture of SDN-native big data integration in our previous paper \cite{alwasel2017programming}. By considering the importance of different policy requirements of big data applications and SDN traffic engineering, BigDataSDNSim is designed to provide mixed composition of big data and SDN policies to be used or extended for evaluating and testing new algorithms and techniques.

\hfill \break

\subsection{Architecture}
\begin{figure}[t] 
	{\includegraphics[]{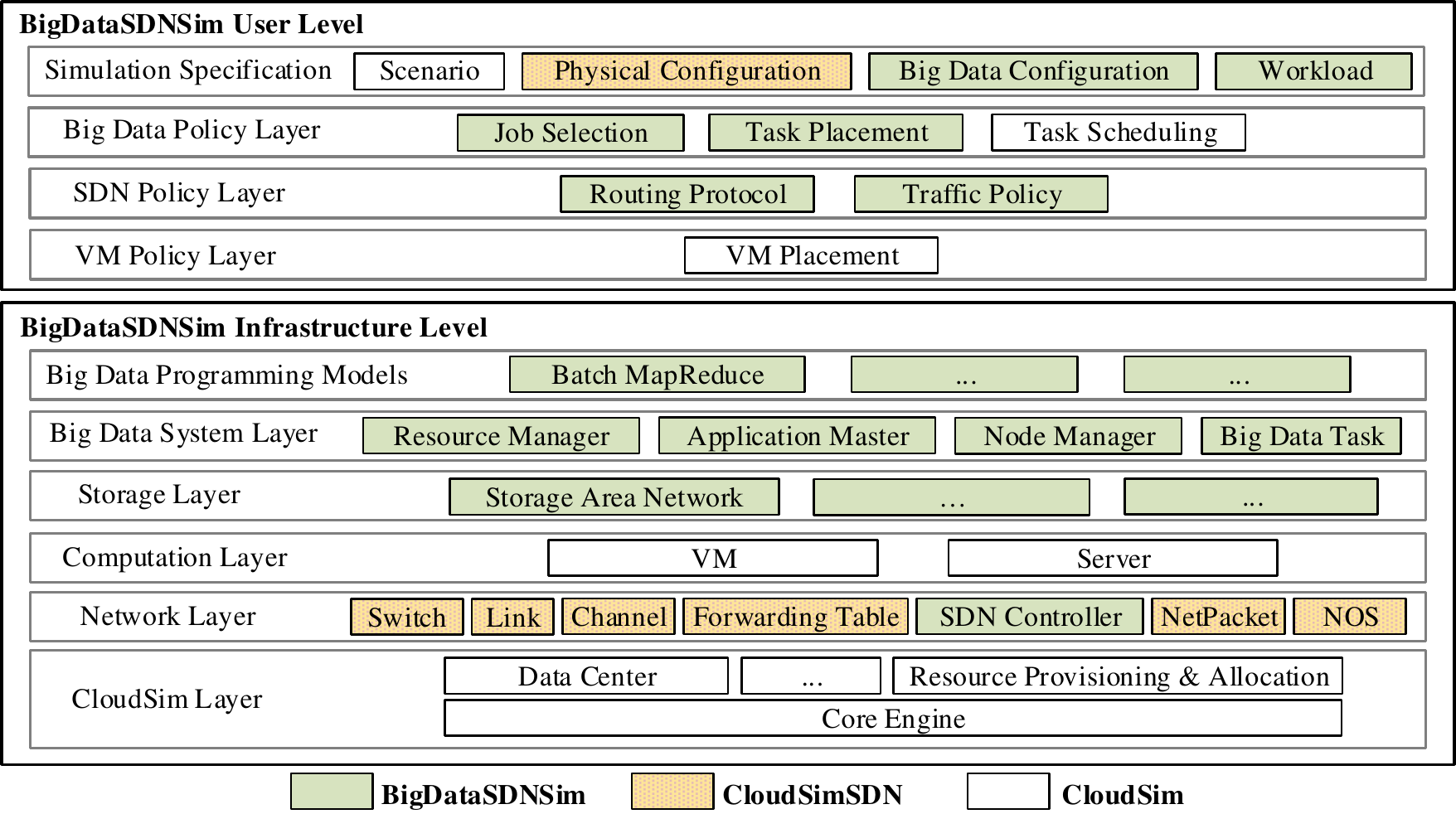}}\label{fig:esssxample}%
	\caption{Architecture of BigDataSDNSim.}%
	\label{fig:BigDataSDNSim_Architecture}%
\end{figure}

Inspired by the designs and architectures of CloudSim, CloudSimSDN, and IoTSim, we developed the BigDataSDNSim simulator to provide the necessary infrastructures for evaluating and analyzing big applications in SDN-powered cloud data centers. Figure \ref{fig:BigDataSDNSim_Architecture} presents key components within a multi-layered architecture of the BigDataSDNSim framework (shown by green boxes), in addition to the used elements of CloudSim and CloudSimSDN. Since IoTSim is specifically developed to simulate the MapReduce model, we modified most of its logic and code to make it applicable to SDN and big data management systems. In later sections, we discuss the design and functionality of our proposed BigDataSDNSim's components in detail.

\subsubsection{User level}
We recommend that most of the tool's users focus on this level to implement their strategies, policies, and algorithms. The user level abstracts away the complexity of the tool, which allows users to focus more effort on cross-performance evaluation and testing of SDN-enabled big data applications. 

\begin{itemize}
\item \textit{Simulation Specification Layer}: This layer allows users to define their own simulation scenarios. They should define the data center descriptions of host resource requirements (e.g. memory, CPUs) and network physical topology configuration (switches' configurations and link relations between hosts and switches) using a single JSON file. Users should also specify big data configurations (the number of required VMs and application masters, the task slot size of application masters, etc.) using a provided Java class. The descriptions of MapReduce workloads or end-users should be submitted in a CSV file, which includes many attributes such as user's ID, job type, start time, and the size of network packets.

\item  \textit{Policy Layers}: They include big data, SDN, and VM policies. Such policies are key factors for obtaining high performance in the MapReduce processing and transmission in cloud data center. The abstract class of VM allocation/placement policy residing in hosts is originally developed in CloudSim. To easily manage MapReduce jobs and tasks, we developed two abstract classes (job selection and task placement) in addition to the use of the task scheduling class developed in CloudSim to support big data management systems with procedures about (i) which job to choose first; (ii) where to place tasks; and (iii) how node managers execute the tasks, respectively. We also developed two SDN abstract classes to allow the SDN controller to dynamically build routing/forwarding tables together with enforcing application traffic policies in the network. More details of the policy abstract classes are discussed in Subsection \ref{sec_biult_in_modules}.

\end{itemize}

\subsubsection{Infrastructure level}
This level contains the core components of our tool. Users who demand more functionalities, big data programming models, and SDN capabilities should extend this infrastructure level. 

\begin{itemize}

\item  \textit{Big Data Programming Models Layer}: Maintaining various programming models of big data is essential to support applications that demand different processing mechanisms. big data applications often require cross-platform task executions; thus, this layer integrates big data programming paradigms under one roof to allow the simulation and analysis of big data applications. Currently, we implement the computing model of MapReduce in SDN-enabled cloud data centers. MapReduce operates on large amounts of data simultaneously in some specific order. Any extra models required by users can be modeled and presented in this layer. 

\item \textit{Big Data Programming System Layer}: Holds components responsible for simulating the behaviors and processes of BDMS, discussed later in Subsection \ref{sec_bdms_sdn_modeling}. BDMS is responsible for allocating and managing resources of distributed systems along with scheduling tasks to numerous types of big data applications running in multiple big data clusters. The main components of BDMS are a resource manager, node managers, application masters, in addition to our generic big data task (BigDataTask) component to simplify the processing of different big data tasks, such as MapReduce, Stream, etc. 

\item  \textit{Storage Layer}: Supports the simulation of various types of storage services that obtain a collection of datasets generated from diverse resources). There are many existing types of storage with different functionalities, processing delays, and network delivery capabilities, such as storage area networks (SAN), network attached storage (NAS), and directly attached storage (DAS). Moreover, big data has its own types of storage infrastructures to store, manage, and forward real time or historical data to cluster nodes, such as Apache Kafka\cite{ApacheKafka} and Hadoop Distributed File System (HDFS\cite{HDFS}). They are critical elements to be considered on simulation scenarios due to their impacts on the overall application performance. Since our work is based on MapReduce, we only consider the modeling of SAN storage and some functionalities of HDFS where historical datasets flow through multiple nodes via the network infrastructure, discussed in later sections. 

\item  \textit{Computation Layer}: Includes physical and virtual components. In clouds, virtualization of resources, especially host virtualization, is the key factor for their success because it offers several benefits, such as on-demand resource allocation, scalability, flexibility, cost-saving, and easy management. With virtualization, hosts are able to share their resources among multiple VMs, where each VM has its own memory, storage, and processor speed characteristics. Note that big data clusters use container-based virtualization to execute applications. However, CloudSimSDN does not have containerized virtualization components yet; therefore, it uses VMs as a replacement for containers. 

\item  \textit{Network Layer}: This layer exposes and simulates network and SDN entities. The fundamental functionalities and issues of SDN deployment and management are handled in this layer. It allows the network to be controlled and managed via one or more SDN controllers fed by routing and traffic policies. It also enables the acquisition of a global network view, network optimization, and the reconfiguration of networks. This layer should be extended when more SDN functionalities are required to reflect realistic experiments. 

\item  \textit{CloudSim Layer}: This layer is equipped with the core entities, functionalities, and engine of CloudSim (data center, resource provisioning and allocation, event processing, simulation clock, etc.). It provides virtualized cloud data centers and it  is responsible for managing the simulation life cycle. The BigDataSDNSim simulator operates on top of this layer where entities communicate via its event engine by extending the SimEntity component. 
\end{itemize}

\begin{figure}[b] 
	{\includegraphics[]{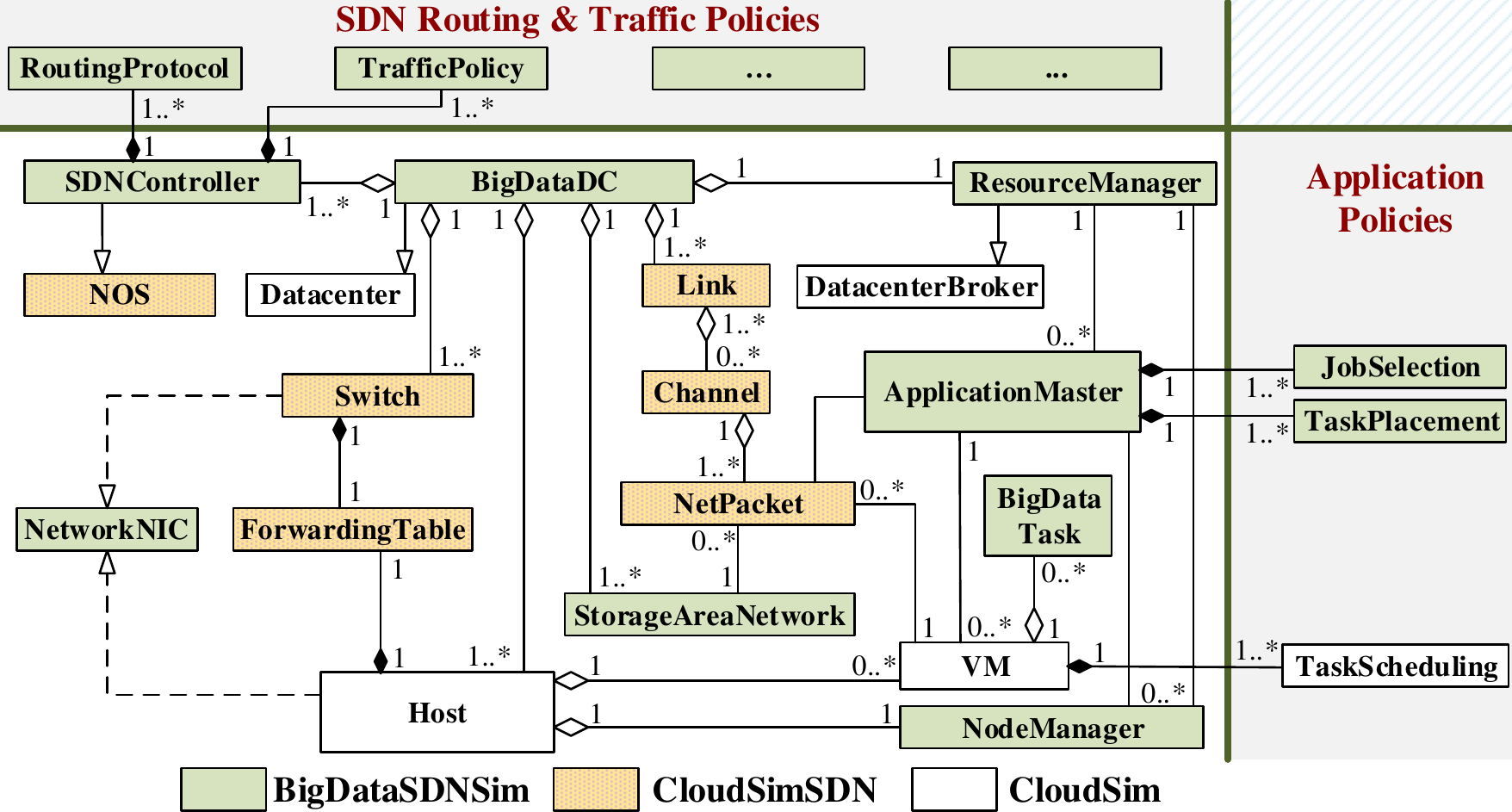}}\label{fig:esssxample}%
	\caption{Class diagram of BigDataSDNSim.}%
	\label{BigDataSDNSim_ClassDiagram}%
\end{figure}

\begin{figure}[t] 
	\centering
	{\includegraphics[]{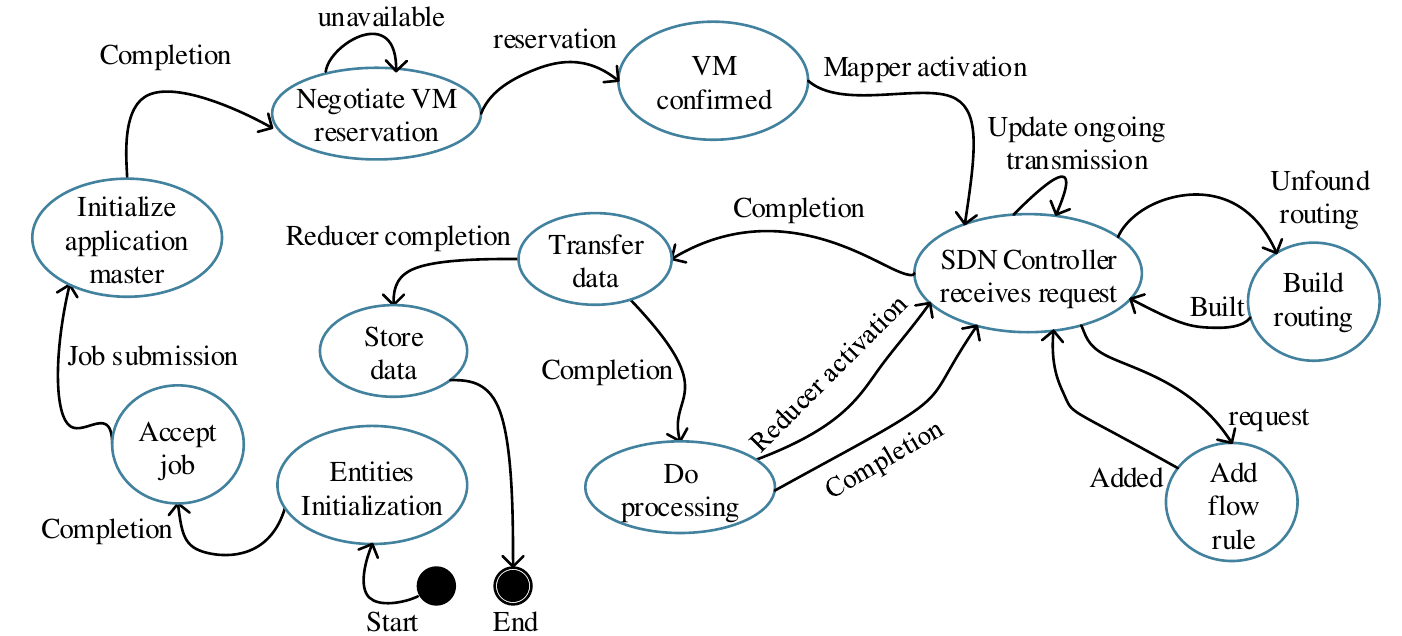}}\label{}%
	\caption{State transition diagram of BigDataSDNSim.}%
	\label{fig_state_transition}%
\end{figure}

It is worth mentioning that VMs, in CloudSim, process applications' tasks using a unit of representation/identification called Cloudlet. Relationsl between Cloudlets and network packets do not exist and thus cannot be identified/linked to each other. The task processing size of VM is different compared to the required network packet's data size to be sent to destination(s); BigDataSDNSim, therefore, uses network packet objects introduced by CloudSimSDN to identify the size of data to be sent over network routes. The size of packet(s) is multiplied by a factor of interests to represent the size of VM tasks (Cloudlets). This facilitates the simulation of task processing mechanisms of big data applications along with network packets that travel inside SDN cloud data centers.

subsection{Design}
In order to implement the functionalities of BigDataSDNSim, we take advantage of the CloudSim's capabilities, mainly SimEntity and SimEvent components. Every element that requires communication and cooperation with other elements (e.g., data center, resource manager, SDN controller, etc.) must extend/use the SimEntity component. The SimEvent component is used to store many attributes, such as IDs of senders and receivers, data directed to destinations, event invoking time, etc. Figure \ref{BigDataSDNSim_ClassDiagram} depicts the main classes of BigDataSDNSim, shown by green boxes. Figure \ref{fig_state_transition} briefly shows a state transition diagram of BigDataSDNSim from start to finish. A given condition must be met to move from one state to another. The following subsections describe the main classes/entities and interfaces used in BigDataSDNSim modeling.     

\hfill \break

\subsubsection{Application modeling design}
To model big data applications that leverage MapReduce programming model for analyzing high volumes of data, we design an BigDataTask class inherited from the Cloudlet class. BigDataTask has numerous variables that need to be filled with values, such as the ID of application and type of task (e.g. Map, Reduce, Stream). It is instantiated by a respective ApplicationMaster based on the number of tasks required for analyzing big data data. The execution of BigDataTask takes place on VMs and is managed by an associated NodeManager residing in a respective host.

\subsubsection{BDMS and SDN modeling design}\label{sec_bdms_sdn_modeling}
To model the  behavior of BDMS within SDN-enabled data centers, the following classes/entities have been designed and added to BigDataSDNSim.

\begin{itemize}
\item  \textit{ResourceManager}: This entity extends the DatacenterBroker class, developed in CloudSim, with more functions and attributes. It is responsible for the configuration, deployment, and scheduling of slave nodes' resources among applications. It continuously waits for users to submit their applications' resource requirements to be provisioned in its cluster. It tries to reserve the required resources using VMs' usage and availability statistics obtained from continuous heartbeat messages of NodeManager daemons. If resources are not enough, it will hold the requests in its resource reservation queue on a first-come first-served basis until the required cluster resources become available. 

\item  \textit{ApplicationMaster}: This is an entity that manages the application life-cycle of giving big data programming models. In order for users to deploy their applications, they must request the ResourceManager to build an ApplicationMaster for each required application. The ApplicationMaster is responsible for negotiating resource reservations, scaling up/down resources, and tracking/monitoring the execution progress of tasks on VMs. A single user can have two big data applications based on a MapReduce programming model with different compositions of code logic and/or multi-level of map and reduce tasks (e.g. one application has two levels of map and reduce tasks, while the other has two levels of map tasks and one level of reduce tasks). Since resources of every VM are shared among different jobs belonging to a same application, the ApplicationMaster should determine scheduling policies for jobs to execute their tasks, such as time-shared, first-come-first-serve, priority, etc.   

\item  \textit{NodeManager}: This class models a node manager, which is in charge of controlling and monitoring host resources. Every node manager is allocated in a single host to periodically report its host status. It consistently informs the actual usage of its host to the ResourceManager, which is used to properly allocate cluster's resources among applications besides alleviating resource contention. 

\item  \textit{NetworkNIC}: Is an interface that must be implemented by every node that needs to establish and maintain a network connection. It is similar to the network interface card (NIC) embedded in most of today's devices. It obtains several abstract functions that should be overridden when necessary to support the simulation of real world Cloud data center networks. The key feature of the interface is to allow the deployment of southbound network management protocols (OpenFlow and Open vSwitch\cite{openvswitch} (OVS)) in switches and hosts of BigDataSDNSim, respectively. Such protocols enable SDN controller(s) to have a full network control of nodes, build routing/forwarding tables, and capture a global network view, to name a few. 

\item  \textit{SDNController}: Is an entity designed to mimic the behavior of an SDN controller. It extends the NetworkOperatingSystem (NOS) class developed in CloudSimSDN with more functions and attributes. It provides network service abstractions for programmability, scalability, and monitoring. Its key responsibility is to dynamically configure and maintain a network's logical topology. By gathering the information from each node (switches and hosts), the controller builds dynamic routing for each host, VM, or application based on implemented routing algorithms. The controller is developed with the capabilities to seamlessly shape network traffic for each application/job by using the northbound interface of traffic policies.

\begin{figure}[t] 
	{\includegraphics[]{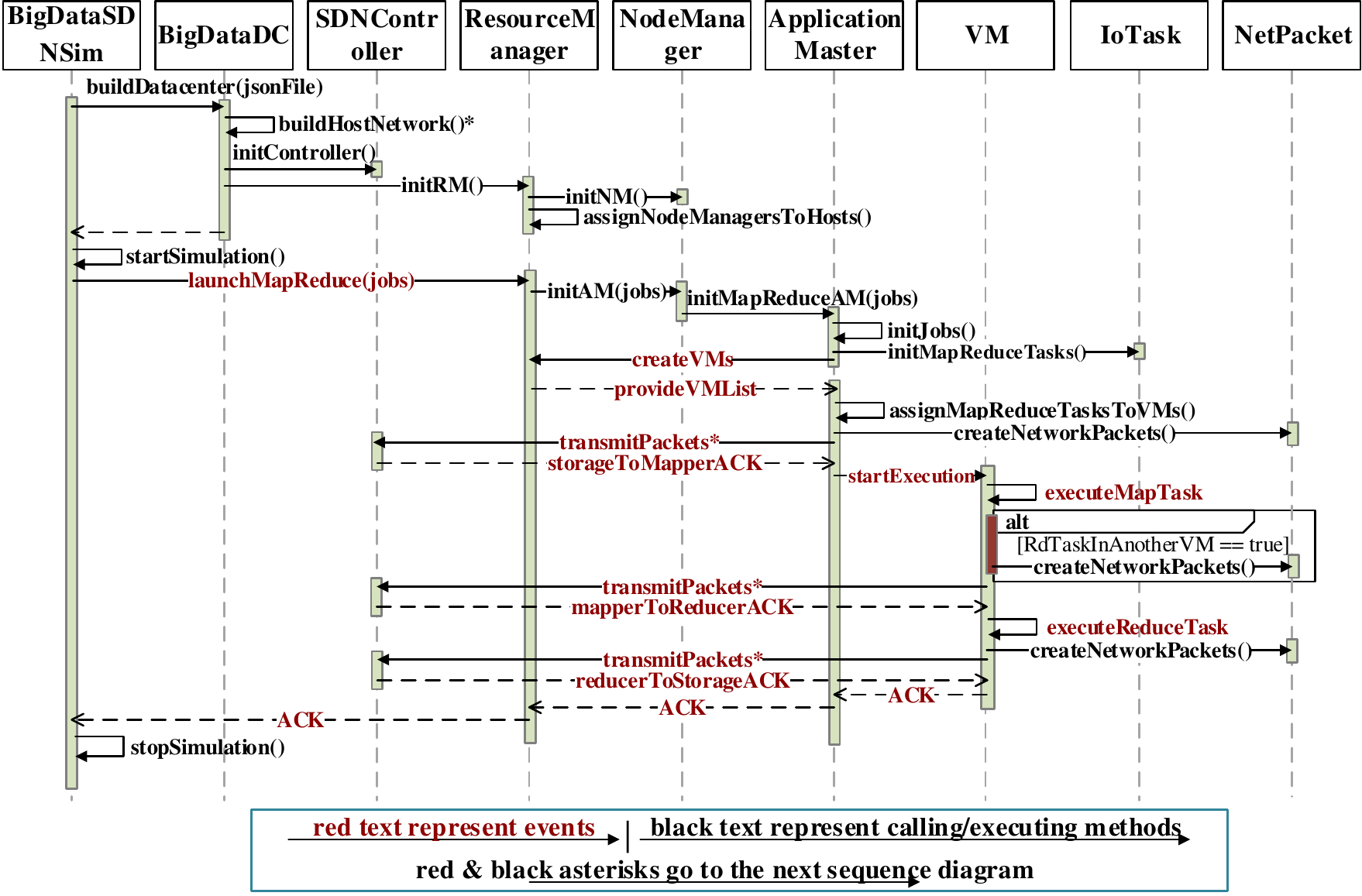}}\label{fig:esssxample}%
	\caption{Sequence diagram of the proposed BDMS, MapReduce, and SDN models.}%
	\label{sequenceBiGData}%
\end{figure}

\begin{figure}[t] 
	{\includegraphics[]{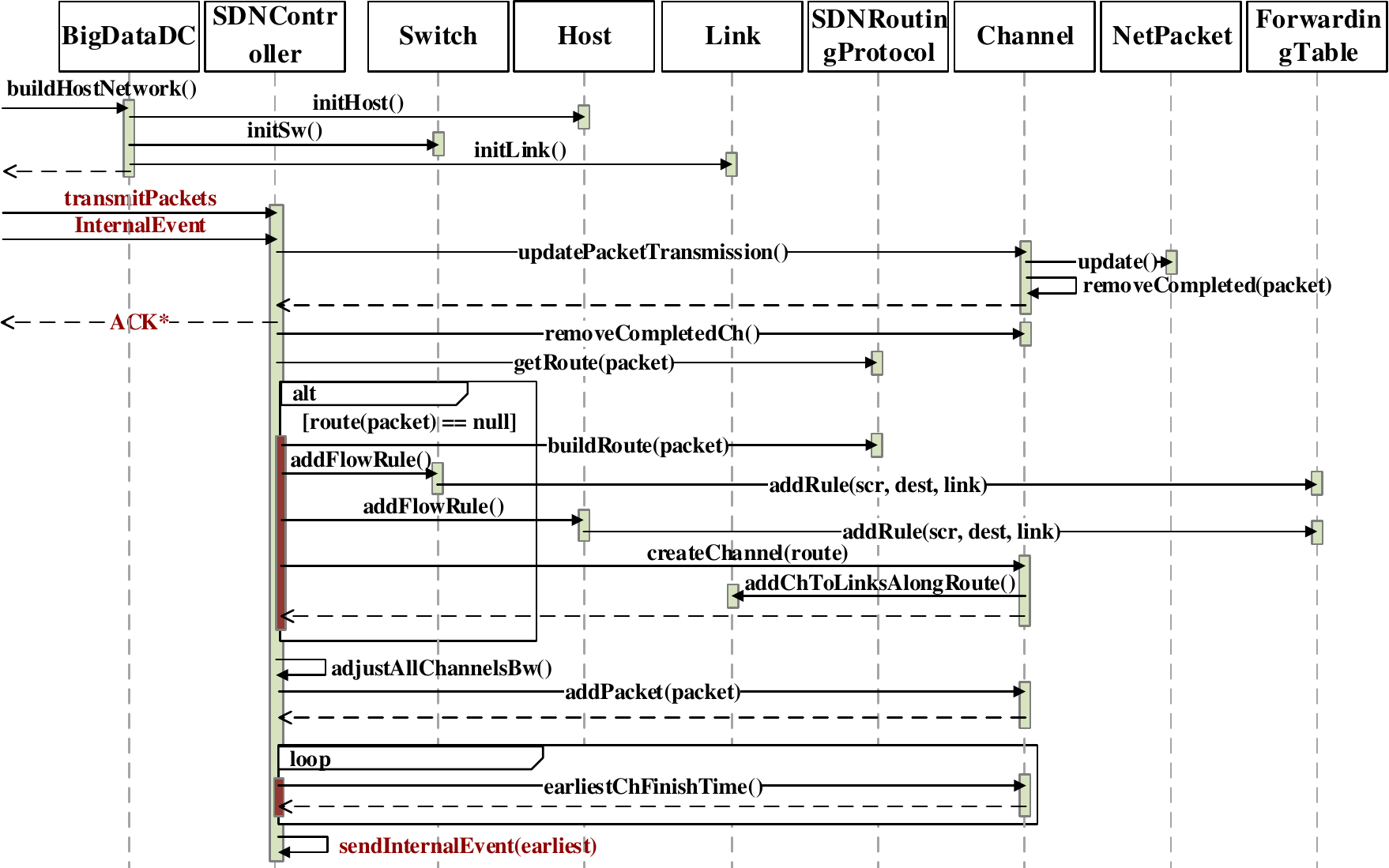}}\label{fig:esssxample}%
	\caption{Sequence diagram of the proposed SDN model in detail (Note: some of the functions derived from CloudSimSDN).}%
	\label{sequenceSDN}%
\end{figure}

\item  \textit{StorageAreaNetwork}: This class represents one type of storage often used in public/private cloud data centers. As a massive volume of data is generated every day from a remarkably large number of applications and devices, storing the data for later analysis and processing is critical, especially when computational resources are unable to handle the incoming stream/batch data. In MapReduce-based applications, data often reside on traditional storage where they leverage distributed file mechanisms (e.g. HFDS) to spread data blocks across elected cluster nodes via network infrastructure. 
\end{itemize}

\subsubsection{Extensions of CloudSim and CloudSimSDN}
Figure \ref{BigDataSDNSim_ClassDiagram} shows the classes of BigDataSDNSim along with the used classes of CloudSim and CloudSimSDN. We implemented a new class called BigDataDC inherited from the Datacenter class of CloudSim to encapsulate all the infrastructure components (networking, computing, and big data systems). Classes of CloudSimSDN, excluding the switch class, are modified with new functions/attributes to enable coordination among big data systems, network infrastructure, and SDN controller(s). Numerous classes and interfaces of CloudSimSDN that are not presented in Figure \ref{BigDataSDNSim_ClassDiagram} (e.g. physicalTopology, virtualTopology, transmission) are removed or moved in the form of methods in other classes to facilitate the implementation and usage of BigDataSDNSim.

\begin{figure}[b] 
	{\includegraphics[]{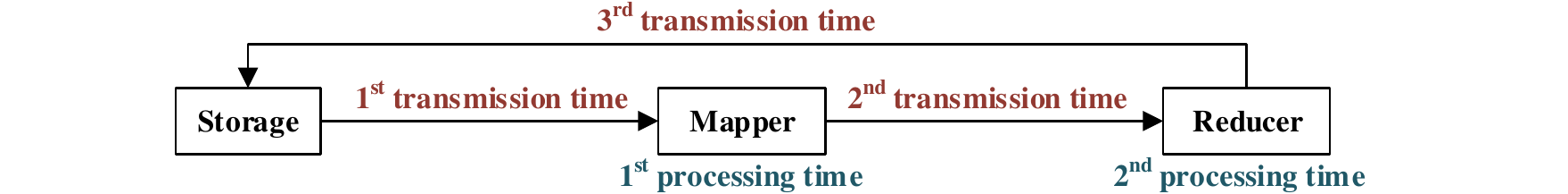}}\label{fig:esssxample}%
	\caption{Transmission and processing phases of BigDataSDNSim's MapReduce model.}%
	\label{fig_mapreduce_workflow}%
\end{figure}

\section{Implementation}\label{sec_implementation}

Figures \ref{sequenceBiGData} and \ref{sequenceSDN} demonstrate the interactions and workflows among classes/entities throughout the runtime of BigDataSDNSim. For simplicity, iterations and other complex data structures used in the simulator are not depicted in the figures. To describe the implementation and function of our simulator as shown in the figures, we classify BigDataSDNSim's lifetime into four phases: infrastructure construction, application establishment, processing and transmission, and performance results.

The infrastructure construction is the first step users must undertake to use the simulator. As described earlier, the configuration of hosts, switches, and links (number of CPUs, size of memory, link's bandwidth etc.) is supplied in a JSON file. Once BigDataSDNSim obtains the file, it passes the file to the BigDataDC by invoking a buildDatacenter() function. As a result, the BigDataDC initiates the corresponding objects of hosts, switches, and links. In addition, it creates the objects of SDN controller and resource manager along with building the network topology. Once the resource manager is active, it couples every physical host with a node manager for monitoring and reporting purposes. Immediately after the successful infrastructure construction, BigDataSDNSim starts the simulation by triggering an embedded function of CloudSim called startSimulation().

Shortly after BigDataSDNSim starts, the application establishment takes place. BigDataSDNSim sends a request in the form of CloudSim's event to the ResourceManager to start the application deployment, which creates a single application master for each requested application. The configuration requirement of jobs (e.g., start time, size of packets, number of node managers, etc.) is provided to the ResourceManager using a CSV file and Java class called BigDataConfigure. Next, the ApplicationMaster iterates over the list of jobs, stores every job and its BigDataTasks  (in the form of map and reduce tasks), and negotiates resource reservation of VMs with the ResourceManager. Note that, we leverage VM-based virtualization as an alternative choice of container-based virtualization due to an unsupported containerization feature in BigDataSDNSim. Once VMs are successfully allocated, theResourceManager sends the list of VMs to the ApplicationMaster.

\begin{figure}[b] 
	{\includegraphics[]{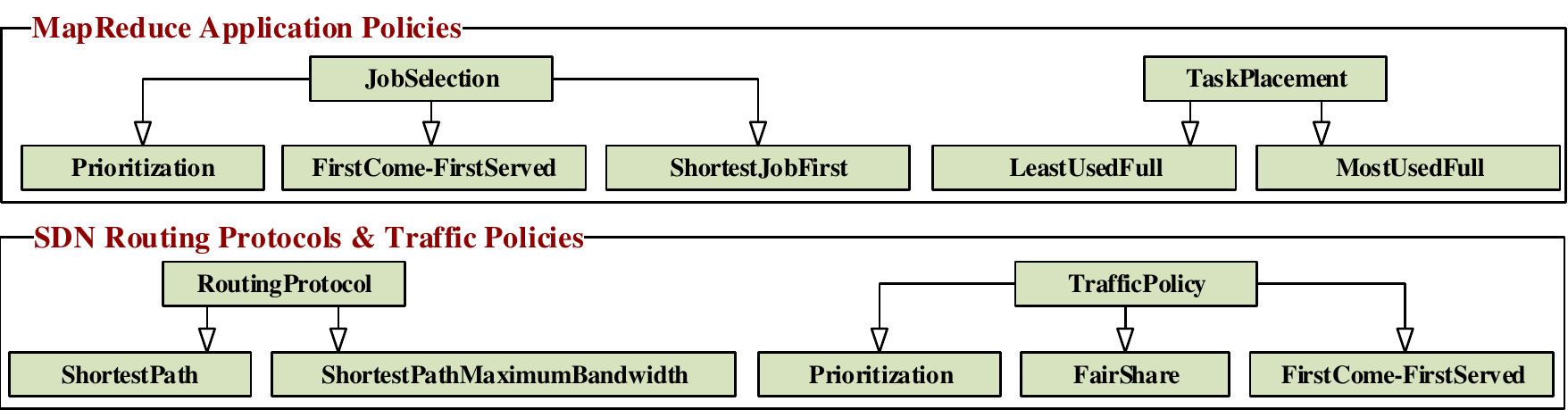}}\label{}%
	\caption{Abstract and extended classes of BigDataSDNSim's policies.}%
	\label{fig_policies}%
\end{figure}

To start the processing and transmission activities, every application should be provided with a list of jobs. Each job is comprised of computational (Cloudlets) and transmission (netPackets) tasks to be executed/transfered in a specific order. The current application logic supported in our simulator is based on two processing and three transmission activities (MapReduce traditional logic), as depicted in Figure \ref{fig_mapreduce_workflow}. For the processing stage to take place, two requirements must be met: (i) the execution logic of map and reduce tasks must be placed on/mapped to VMs and (ii) mappers and reducers must acquire the whole required data to start execution. The transmission stage, on the other hand, includes three activities (i) transferring data from the SAN storage to mappers; (ii) transferring mappers' output (intermediate data) to reducers; and (iii) transferring reducers' output (final data) to the SAN storage, respectively (see Figure \ref{fig_mapreduce_workflow}). The ApplicationMaster calculates the size of data blocks to be transferred from SAN to mappers by dividing a given job data size over the number of mappers. The following equation is used to compute mapper size \textit{ms}

\begin{equation}
 ms = \dfrac{jl}{nm}
\end{equation}

where $jl$ is total job length and \textit{nm} is the number of job's mappers. The data block size of the reducer(s) is subject to the output of its mapper(s). The reducer size \textit{rs} is obtained through the following equation

\begin{equation}
 rs = ms * f
\end{equation}

where \textit{f} is a factor of interest. 

Any element (mapper, reducer, ApplicationMaster, etc.) that requires data transmission must create an object of network packet and send an event (transmitPacket) loaded with the packet to the SDN controller. Once the SDN controller receives the event, it first updates the progress of existing packets and removes completed packets along with idle channels. Next, the controller builds paths/forwarding tables based on VM-to-VM or task-to-task communication (e.g. the SAN storage can send data to map tasks residing in a same VM using different paths), with respect to the implemented SDN routing algorithms. If there is no existing route between the source and destination, then the SDN controller will build a route, add a forwarding rule to every node along the route (switches and hosts), and create a channel. The channel is encapsulated in every link along the route in order to avoid a network bottleneck (e.g. the volume of transmitted data is higher than the bandwidths of intermediate nodes). Following that, the SDN controller adjusts the bandwidth of every channel by finding the smallest available bandwidth among physical links existing along the route. The following equation is used to compute channel bandwidth \textit{c\textsubscript{bw}}  

\begin{equation}
 c_{bw} = \dfrac{min(l_{bw}(i))}{nc} 
\end{equation}

where \textit{l\textsubscript{bw}(i)} is the \textit{i}th link bandwidth that a given channel traverses through, and \textit{nc} is the number of channels traveling and sharing the \textit{i}th link. Once the bandwidth is determined, the SDN controller hands over the packet to the intended channel. In order for the SDN controller to track the completion of packets, it calculates the earliest finish time  \textit{{e\textsubscript{ft}}} among packets, creates an event with an invoking time of  \textit{{e\textsubscript{ft}}}, and sends the event internally to itself. The \textit{{e\textsubscript{ft}}} can be computed using the following equation

\begin{equation}
e_{ft} = min(\dfrac{p_{r}(j)}{c_{bw}(j)})  
\end{equation}

where \textit{p\textsubscript{r}(j)} is the remaining data of the \textit{j}th packet to be transferred and \textit{c\textsubscript{bw}(j)} is the channel of the \textit{j}th packet. The estimated transmission time  \textit{tr\textsubscript{t}(j)} of the \textit{j}th packet can be determined using the following equation

\begin{equation}
tr_{t}(j) = \dfrac{d_{s}(j)}{c_{bw}(j)}  
\end{equation}

where \textit{d\textsubscript{s}(j)} is data size of the \textit{j}th packet to be transferred. The job's total transmission time \textit{j\textsubscript{tr}} is given by 

\begin{equation}
j_{tr} = max(s_{tr}(sm)) + max(mp_{tr}(mr)) + max(rd_{tr}(rs))
\end{equation}

where \textit{s\textsubscript{tr}(sm)} is transmission time from storage \textit{s} to mapper \textit{m}, \textit{mp\textsubscript{tr}(mr)} is transmission time from mapper \textit{m} to reducer \textit{r}, and \textit{rd\textsubscript{tr}(rs)} is transmission time from reducer \textit{r} to storage \textit{s}. The following equation is used to compute the execution time of mappers for a given job

\begin{equation}
j_{mp} = max(mp_{e}(m))
\end{equation}

where \textit{mp\textsubscript{e}(m)} is the \textit{m}th mapper execution time . The following equation is used to compute the execution time of reducers for a given job

\begin{equation}
j_{rd} = max(rd_{e}(n)) 
\end{equation}

where \textit{rd\textsubscript{e}(n)} is the \textit{n}th reducer execution time. The job completion time \textit{j\textsubscript{ct}} is given by 

\begin{equation}
j_{ct} = j_{tr} + j_{mp} +  j_{rd} 
\end{equation}

Once there is no more activity/event to take place, the simulation concludes and results are reported. The report structure reflects the information of jobs, performance measurement of transmission and processing, energy statistics of network and hosts, and information of forwarding tables. The output of jobs presents the status of every job, such as ID, submission time, queuing delay, start time, etc. The transmission result demonstrates the statistics of every connection (IDs of source and destination, size of packets, start time, transmission time, etc.), while the processing result shows the performance of each map/reduce task (e.g. VM's ID, start time, execution time). The energy result presents the power consumption of every switch and host. The information of forwarding tables shows the list of traversing nodes for every packet along with changes made to the packet's bandwidth (size and time) throughout the transmission period.

\subsection{Built-in module policies}\label{sec_biult_in_modules}
Computation and communication resource management play an important role in application performance. Their improper management can lead to numerous issues, such as performance degradation, increasing costs, and wasting energy. Therefore, we broke down the management of BDMS and SDN into four abstract classes. Such abstractions provide flexibility and facilitate the deployment process of computation and communication policies (see Figure \ref{fig_policies}). Those classes should be extended by developers/researchers who aim to deploy their custom policies. For the purpose of evaluating BigDataSDNSim, we extended every abstract class with a few algorithms (e.g, shortest path first, least used first), discussed in the next section.

\begin{figure}[b] 
	{\includegraphics[]{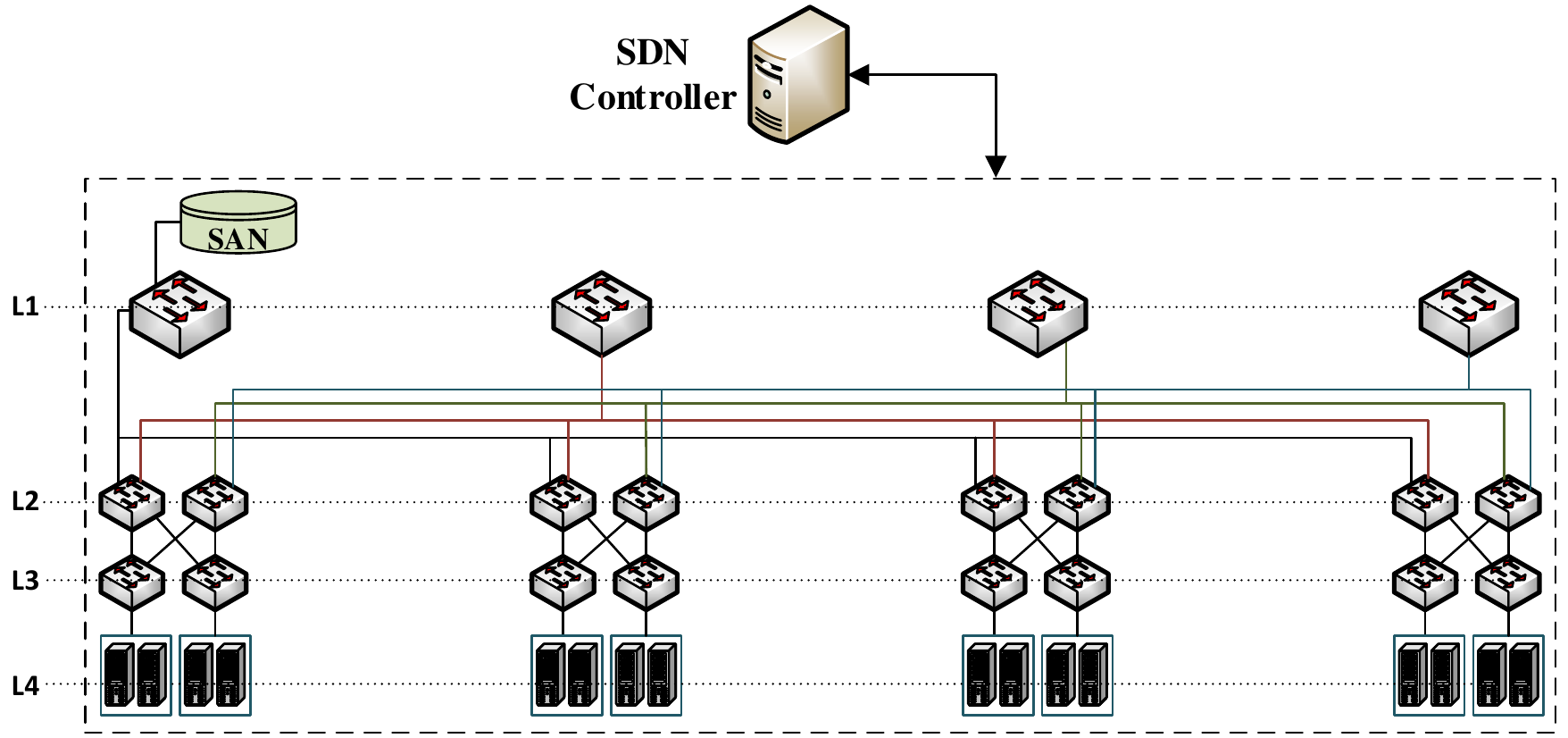}}\label{fig:esssxample}%
	\caption{Architecture of the BigDataSDNSim simulation environment with three-tier SDN-enabled Cloud data center.}%
	\label{fig_network_physical_configuration}%
\end{figure}

\begin{itemize}
\item  \textit{Job selection and task placement}: The job selection and task placement plays an important role in the overall job performance. Every application master has a job scheduler to control the execution of jobs. The application master queues all incoming jobs and schedules them based on a given selection policy and resource availability. Once a job is ready to be processed, the application master reads the job's task requirements (e.g. number of mappers, number of reducers) and places the tasks into elected VMs based on a given task placement policy. The tasks do not take place until they are triggered. Note that, the VM's task scheduling is already implemented in CloudSim with two scheduling techniques (time-shared and space-shared).

\item  \textit{Routing protocol and traffic policy}: As described earlier, the SDN paradigm abstracts away embedded routing mechanisms (a.k.a., forwarding plane) from networking devices (switches and routers) to an SDN controller, refer to Figure \ref{fig_SDN_VS_Legacy}. Such abstraction requires the SDN controller to maintain a whole network picture of all connected devices (physical topology) in order to make proper networking decisions. To this end, the SDN controller must contain several protocols to discover devices, maintain pairwise relations between devices, build forwarding rules, manipulate devices' forwarding tables, etc. For the SDN controller to be able to dynamically build routing/forwarding tables, we implemented an abstract class (RoutingProtocol) that can be extended with custom SDN routing algorithms. We adopted and implemented Dijkstra's algorithm \cite{dijkstra1959note} in a fat-tree topology, which is capable of finding the shortest paths from a single source (e.g. VM, mapper, reducer) to all other destinations based on given objectives (minimum number of traversing switches,  maximum bandwidth of a route, etc.). Moreover, network traffic can be shaped based on users' QoS requirements, such as prioritizing applications/jobs to meet deadline constraints. We, therefore, implemented another abstract class (TrafficPolicy) that can be extended with novel SDN traffic policies, discussed in Subsection \ref{sec_sdn_mr_policies}.  

\end{itemize}

\section{Performance Evaluation}\label{sec_evaluation}

In order to demonstrate the practicality of BigDataSDNSim, we created a use-case based on a MapReduce application. The use-case shows a performance and energy comparison of legacy networks and an SDN paradigm in cloud data centers. The transition from traditional networks to the SDN is still in its infancy and several investigations need to be made. We therefore shed light on some of SDN's features by showing how SDN-enabled cloud data centers are capable of not only optimizing application performance but also of reducing energy consumption.

\begin{table}[t]
  \centering
  \caption{Data center configuration}
    \begin{tabular}[t]{|p{2cm}|l|r|}  
     \hline
	\multicolumn{2}{|c|}{\textbf{Host \& SAN}} \\
     \hline
    CPUs & 8 \\
 \hline
    RAM (GB) & 30 \\
     \hline
    MIPS & 10000 \\
 \hline
    \end{tabular}%
 \hfill
  \centering
    \begin{tabular}[t]{|p{2cm}|l|l|}
    \hline
     \multicolumn{2}{|c|}{\textbf{VM}} \\
    \hline
    CPUs & \multicolumn{1}{r|}{4} \\ \hline
    RAM (GB) & \multicolumn{1}{r|}{8} \\ \hline
    MIPS & \multicolumn{1}{r|}{1250} \\ \hline       
    \end{tabular}%
 \hfill
  \centering
    \begin{tabular}[t]{|p{5cm}|l|l|r|}
       \hline
    \multicolumn{2}{|c|}{\textbf{Network}} \\
   \hline
        Link & Bandwidth \\ \hline
    Storage <-> Core1 switch & 4 Gbps \\ \hline
    Core <-> Aggregate switches & 1 Gbps \\ \hline
    Aggregate <-> Edge switches & 1 Gbps \\ \hline
    \end{tabular}%
  \label{datacenter_configuration}%
\end{table}%

\subsection{Experiment Configurations}\label{subsection_network_config}
We created a single cloud data center with a hierarchical network architecture (three-tier layers). Figure \ref{fig_network_physical_configuration} depicts the physical topology that includes three layers of switches and one layer of hosts. There are four core switches, eight aggregation switches, eight edge switches, and 16 hosts. Physical nodes (hosts and switches) and virtual machines are configured according to Table \ref{datacenter_configuration}. There is a single link coupled with a bandwidth of 4 Gbps between the SAN storage and one of the core switches (core1). Starting from the left side of the figure, the first pair of core switches (L1) is connected to four odd switches of the child layer (L2) by two links, configured with a bandwidth of 1 Gbps each, and vice versa to the others. Every aggregation switch (L2) is attached to two associated edge switches (L3) via a single link each with a bandwidth of 1 Gbps. Similarly, every edge is connected to two hosts, each with a link configured with a bandwidth of 1 Gbps.

\begin{figure}[t] 
	\subfloat[Application establishment]{\includegraphics[]{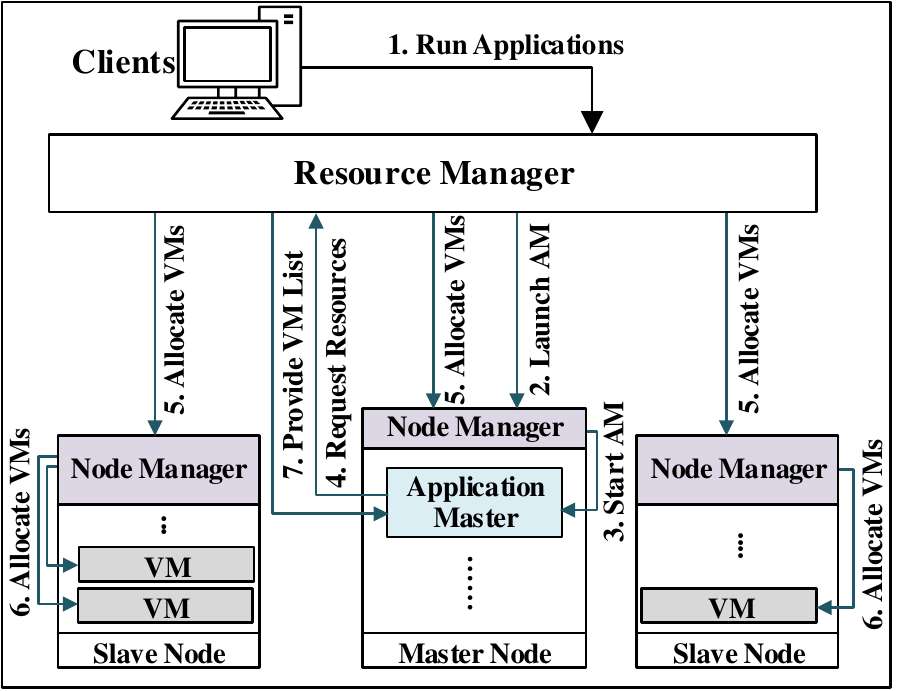}\hspace{0em}}\hfill
	\subfloat[Application processing]{\includegraphics[]{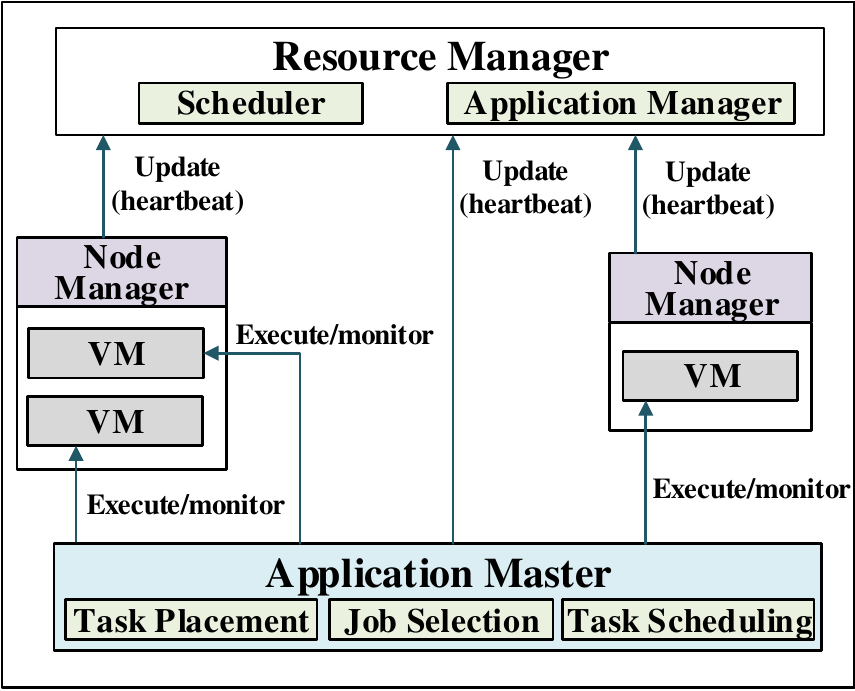}\hspace{0em}}\hfill%
	\caption{Major steps of big data application establishment and processing within the cluster of a big data management system.}%
	\label{fig_app_processing}%
\end{figure}

Hosts and SAN storage are considered homogeneous, as defined in Table \ref{datacenter_configuration}. Every host is attached to a single node manager to manage resources and report utilization to a resource manager (see Figure \ref{fig_app_processing}). As hosts can always be active in real Cloud data centers, the hosts (in the run time of BigDataSDNSim) are always active and wait for a VM allocation/deallocation request from users/customers. Physical storage can be similar to hosts in terms of resource capabilities (CPU, memory, etc.); therefore, we configure the SAN with similar capabilities as hosts. Note that the performance of SAN (read, write, etc.) does not play an important part in the use-case since we only consider the performance of MapReduce applications from the time that data travel from the SAN to mappers until the SAN receives the final data output from reducers, as described in Section \ref{sec_implementation}. 

\begin{table}[b]
  \centering
  \caption{Job configuration}
    \begin{tabular}{|c|c|c|c|c|c|r|r|}
      \hline
    \multirow{2}[4]{*}{\textbf{Job type}} & \multicolumn{2}{c|}
    {\textbf{Date processing  size}} & \multicolumn{3}{c|}{\textbf{Data transmission size}} & \multicolumn{1}{c|}{\multirow{2}[4]{*}{\textbf{Map number}}} & \multicolumn{1}{c|}{\multirow{2}[4]{*}{\textbf{Reduce number}}} \\
    
\cmidrule{2-6} & Mappers & Reducers & Storage & Mappers & Reducers &     &  \\
     \hline
    Small & 100000 (MI) & 75000 (MI) & 200 (Gb) & 150 (Gb) & 100 (Gb) & 2   & 1 \\
       \hline
    Medium & 200000 (MI) & 175000 (MI) & 400 (Gb) & 350 (Gb) & 300 (Gb) & 4   & 2 \\
     \hline
    Big & 300000 (MI) & 275000 (MI) & 600 (Gb) & 550 (Gb) & 500 (Gb) & 6   & 3 \\
   \hline
    \end{tabular}%
  \label{job_configuration}%
\end{table}%

In our use-case, we assume that the transportation and logistics industry has 15 different sets of data (jobs) to be processed with different execution logic of MapReduce. In real-world examples of MapReduce applications, every single or group of mappers and reducers have their own codes/functions where each one processes data in the form of key-value pairs. With the help of BDMS, the industry can have a single application master capable of handling different sub-application logic in the form of jobs. There is only a single resource manager that is always up and running to handle new resource allocation requests of a respective cluster, as shown in Figure \ref{fig_app_processing}.

The configuration of jobs the industry submits is shown in Table \ref{job_configuration}. The jobs are classified into three categories: small, medium, and big. Every category has five jobs where each category differs from one another in data processing size, data transmission size, number of mappers, and number of reducers. Further, the industry requires 16 VMs managed by a single application master. Note that mappers and reducers are processed based on million instructions (MI). The data center is configured to allocate VMs based on a simple VM allocation policy; this has no impact on our experimental evaluations.

 \begin{figure}[t] 
   \subfloat[\label{fig_NetworkTransmission} Comparison of network transmission time]{\includegraphics[width=0.49\linewidth,height=4cm]	{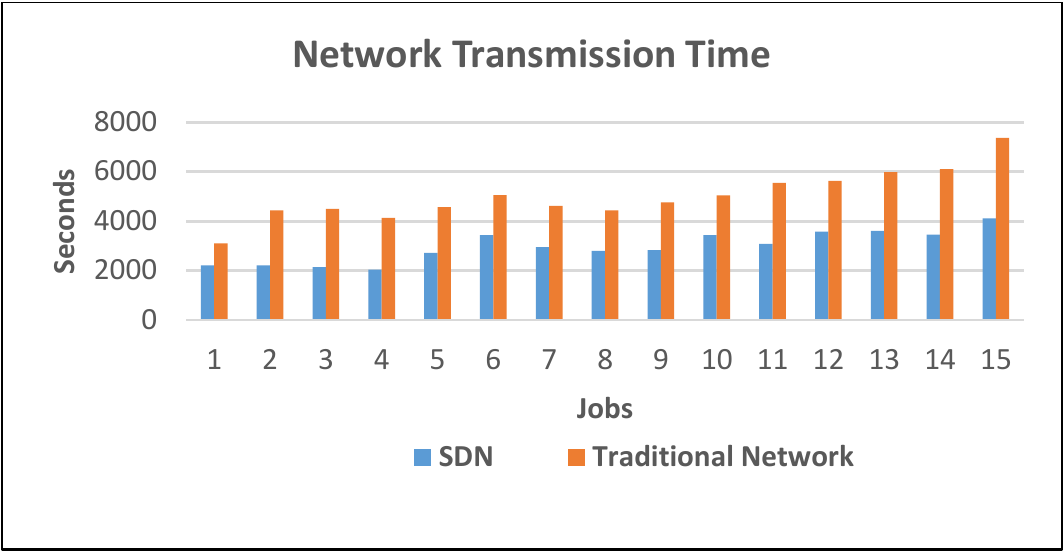}\hspace{1em}}\hfill
   	\subfloat[\label{fig_JobCompletion} Comparison of job completion time]{\includegraphics[width=0.49\linewidth,height=4cm]{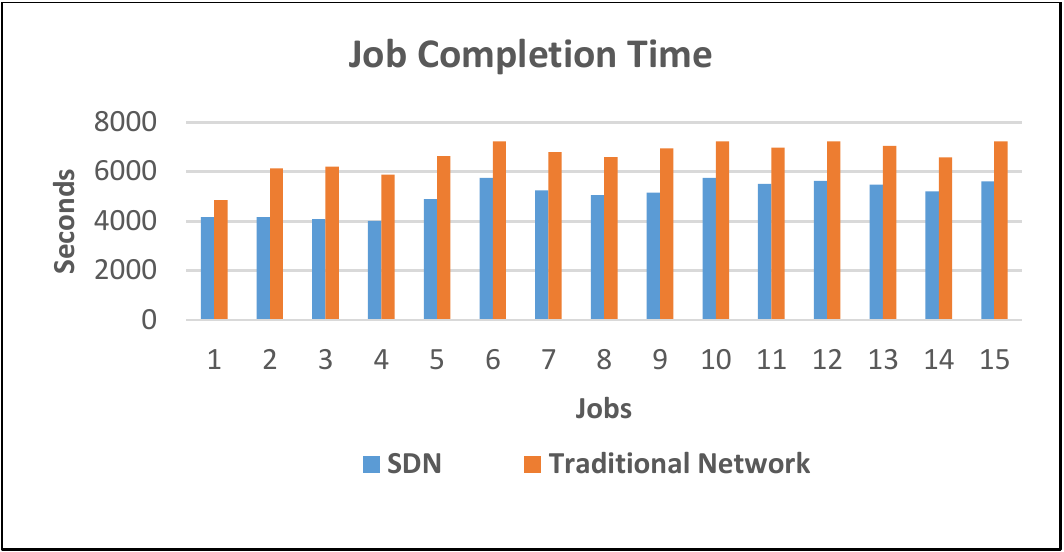}\hspace{1em}}\hfill
	\caption{Comparison between SDN-enabled and traditional networks.}\label{fig_Transmission_Completion}	
\end{figure}

\begin{figure}[t] 
   \subfloat[Comparison of mappers execution time]{\includegraphics[width=0.49\linewidth,height=4cm]{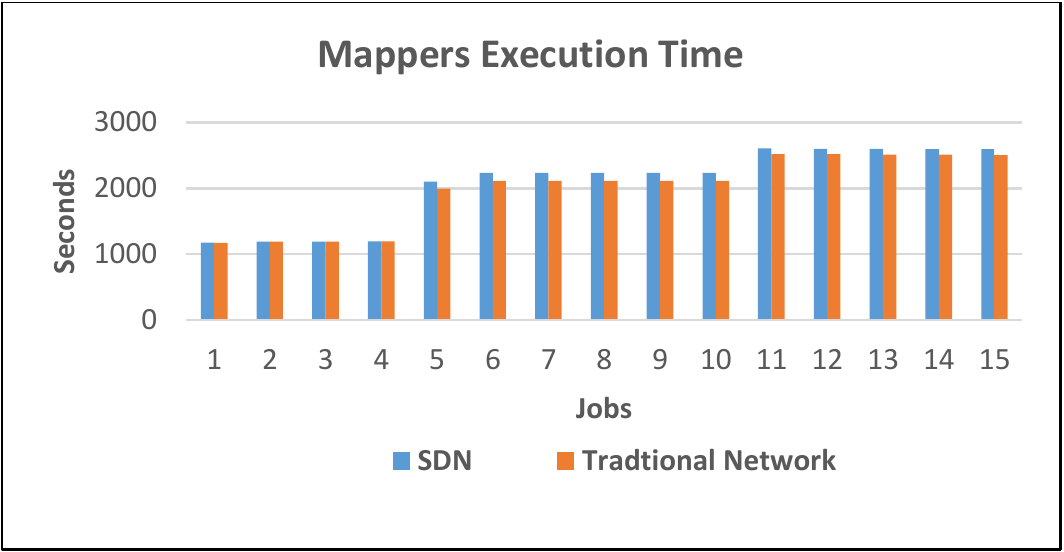}\hspace{1em}}\hfill
   	\subfloat[Comparison of reducers execution time]{\includegraphics[width=0.49\linewidth,height=4cm]{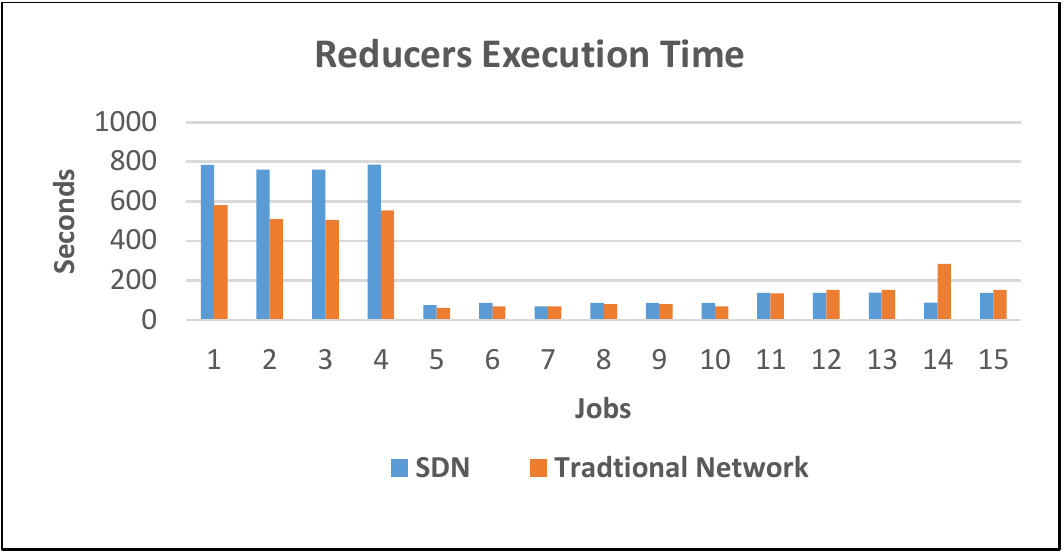}\hspace{1em}}\hfill
	\caption{Comparison between SDN-enabled and traditional networks.}%
	\label{fig_mappers_reducers}%
\end{figure}

\subsection{SDN and application policies}\label{sec_sdn_mr_policies}
In the simulation environment, several policies must be considered throughout the establishment and processing phases of the application along with SDN polices (see Figure \ref{fig_policies}). As mentioned in Section \ref{sec_biult_in_modules}, application-level policies consist of job selection, MapReduce task placement, and VM's task scheduling, while the SDN-level policies contain SDN routing protocol and SDN traffic management. The application-level policies that are considered in the use-case evaluation are (i) a first-come, first-served for the job selection; (ii) VM least-used for the task placement; and (iii) time-shared task scheduling for VMs.
	
    We implemented two routing algorithms to show the difference between legacy networks and SDN along with obtaining dynamic routing and traffic shaping capabilities. For legacy networks, we implemented the well-known Dijkstra's algorithm with the objective of finding the minimum number of traversing links (shortest path). Note that we use a fat-tree topology where there are many routes with same number of links. The algorithm finds all elected routes and randomly selects one of them where packets of respective mappers/reducers will only travel via the route. For SDN, we extended the Dijkstra's algorithm with the objective of finding the minimum number of traversing links first, then finding the maximum bandwidth among the elected routes. Every time a new packet enters the network, the SDN controller executes the algorithm and finds a route, constrained to the two objectives. Two packets from the same VM can have two different routes to the same destination VM. Moreover, we used a fair-share traffic policy where network bandwidth is equally shared.

\subsection{Evaluation Results}
We created the simulation environment and job configurations according to the criteria discussed above. The order of jobs (small, medium, and large) entering the simulation is random. There is an interval time of one second. The use-case follows the processing and transmission phases depicted in Figure \ref{fig_mapreduce_workflow}. We ran BigDataSDNSim twice: one SDN-enabled and the other with traditional network enabled. We collected the simulating data and made comparisons involving network transmission time, job completion time, mappers execution time, reducers execution time, and energy consumption. We arranged jobs from smallest to largest in both Figure \ref{fig_Transmission_Completion} and Figure \ref{fig_mappers_reducers} (1 to 5 are small jobs, 6 to 10 are medium jobs, and 11 to 15 are big jobs). From the figures, it is apparent that the SDN paradigm outperforms legacy networks in terms of performance improvement for MapReduce applications as well as energy reduction in cloud data centers. Note that the performance and energy consumption of SAN storage are not considered in the evaluation. 

The number of transmissions a MapReduce model requires during its life cycle is illustrated in Figure \ref{fig_mapreduce_workflow}. The comparison of network transmission performance between SDN-enabled and traditional networks is depicted in Figure \ref{fig_NetworkTransmission}. Results show that the SDN-enabled network transfers data between elements (storage, mappers, and reducers) faster than the traditional network for all jobs, in which MapReduce transmission performance is drastically enhanced by 41\% on average. The reason behind SDN superiority is that the SDN controller allows two or more packets from a single VM to travel to another destination VM via two or more paths where all paths have the same number of links but different bandwidths, as described earlier. 

Figure \ref{fig_JobCompletion} shows the completion time of jobs in both SDN-enabled and traditional networks. The completion time includes all transmission and processing phases starting with data transmission from the SAN storage and ending with data transmission from reducers (see Figure \ref{fig_mapreduce_workflow}). Figure \ref{fig_JobCompletion} shows that with the SDN-enabled data center, every job takes less time compared with the traditional network. SDN improves job completion time by approximately 24\% on average. Every job has a different completion time even though it shares the same configuration with other jobs belonging to the same group. This is because the location of map and reduce tasks of every job is placed on least used  VMs. Such placement would most likely put mappers and reducers in VMs that reside in different hosts, resulting in more time for completion due to the high dependency on the network layer to transfer data to recipients. Further, every network route has a different traffic load and every VM has a different processing load; thus, mappers/reducers belonging to the same job could have different execution and transmission times. 

Figure \ref{fig_mappers_reducers} illustrates the execution time of mappers and reducers. In both SDN and traditional cases, the execution time period of mappers is slightly similar because they start processing at almost the same time. Before mappers start processing their tasks, they need to obtain data from the SAN storage. According to the given use-case scenario, the storage starts transferring data to all mappers at almost the same time. As mentioned earlier, the bandwidth of every packet is subject to the smallest available bandwidth of a link that exists on its route to avoid network bottlenecks. Thus, the bandwidth availability for all packets from the storage to mappers in both networks is almost the same; consequently, the transmission time from the SAN storage to the mappers is roughly similar. Furthermore, as the mappers belonging to the same group share the VMs' resources on a time-shared basis, they would finish their processing with a similar time frame. However, some identical jobs might have different mappers' execution time due to load variance of slave VMs, such as job number five compared with its similar job type (small).

On the other hand, some reducers in the traditional network have shorter execution time in comparison to the SDN-enabled network. This is because the transmission of data to the reducers takes a longer time compared with the SDN-enabled network, giving the reducers' VMs enough time to finish processing other tasks. Therefore, by the time the reducers receive their whole required data, their VMs would be free or slightly used, which results in faster processing times. Conversely, the SDN network transfers data to reducers at a fast speed leading to high processing loads on reducers' VMs, especially when several reducers residing on a single VM wait for data to start executing. Nevertheless, the execution time of reducers has a slight impact on the overall job completion time because our use-case scenario is focused on data-intensive applications more than compute-intensive ones, meaning that network performance is more important than VM performance. Still, there is a need to develop joint-optimization of networking and computing algorithms that coordinate an SDN controller with application masters.

\begin{figure}[t] 
\centering
   	{\includegraphics[width=0.49\linewidth,height=4cm]{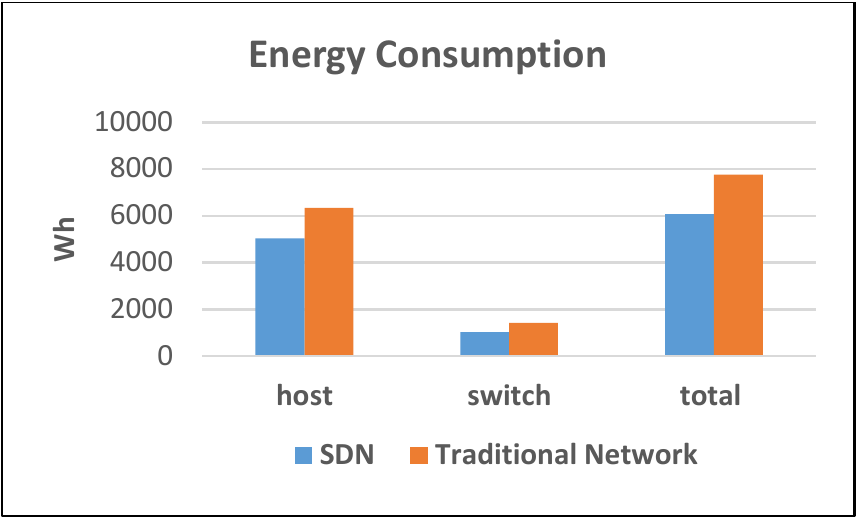}}\hfill
	\caption{Comparison of power consumption between SDN-enabled and traditional networks running MapReduce-enabled applications in Cloud data center.}%
	\label{fig_EnergyConsumption}%
\end{figure}

Figure \ref{fig_EnergyConsumption} depicts the energy consumption of hosts and switches in both SDN-enabled and traditional networks. It can be noticed that the former decreases the power-consumption of its data center by approximately 22\% as against the traditional network. The explanation for this is that the faster the network transfers data, the less power the data center utilizes assuming that idle-mode of hosts and switches is activated. The real-time dynamic routing of SDN is not the only factor that contributes to energy saving, there are also many other factors which we have not considered in depth, such as VM placement/consolidation, MapReduce job placement, traffic consolidation, etc.

Our simple use-case shows how BigDataSDNSim can be used to model and simulate different kinds of cross-layer performance and energy-conscious techniques/policies. BigDataSDNSim could serve as a base for exploring and studying hypotheses and proposed solutions. With the help of BigDataSDNSim, individuals (e.g. researchers, developers) can efficiently evaluate and observe the behavior of their techniques and system design choices. By obtaining the experiment results of BigDataSDNSim, individuals can easily identify and address hidden issues along with building a high level of performance confidence of their approaches and algorithms. 

\hfill \break

\section{Conclusions and Future Work}\label{sec_conclusion}

To enhance the speed of data extraction and analysis, there is an considerable demands for big data applications to coordinate with Software-Defined Networking systems in cloud data centers. However, the fabric and harmony of big data applications and Software-Defined Networking in cloud environments is still in its infancy where they pose several challenges that should be addressed prior to real deployment. The most preferable option for evaluating proposed solutions is  is the use of modeling and simulation-based approaches because of their noticeable advantages (e.g. cost and time saving, flexibility, reproducibility) compared with real cloud testing and evaluation. To meet these demands and requirements, we in this paper presented BigDataSDNSim, a novel simulation-based tool that is capable of modeling and simulating big data applications in SDN-enabled cloud data centers. We presented the design and characteristics of our framework, described its main components, and demonstrated the interrelationships between the components in detail.

BigDataSDNSim is capable of evaluating and analyzing big data applications that utilize MapReduce programming models in SDN-enabled cloud data centers. BigDataSDNSim provides an infrastructure for researchers to quantify the performance impacts of big data applications by means of network and application design choices and scheduling policies coupled with other configuration factors.

We demonstrated the practicality of our framework by presenting a use-case. We studied how multi-level scheduling policies affect the performance of big data applications in SDN-powered cloud data centers. We showed the performance and energy impacts of SDN-enabled versus legacy networks on big data applications in cloud data centers. Our use-case results confirm that an SDN-enabled cloud data center outperforms legacy networks, in which the SDN improves the average of network transmission time by 41\% and the average of job completion time by 24\% as well as reducing the power consumption by approximately 22\% . As our future work, we will focus on the modeling and implementation of a Stream programming model along with clustering-based SDN controllers as well as establishing the correctness and accuracy of BigDataSDNSim using a real environment.

\section*{Acknowledgments}
The work in this paper is supported by Saudi Electronic University (SEU) through the Saudi Arabian Culture Bureau (SACB) in the United Kingdom. This research is also supported by three UK projects LANDSLIP: NE/P000681/1, FloodPrep:NE/P017134/1, and PACE: EP/R033293/1.

\nocite{*}

\bibliography{wileyNJD-AMA}%

\begin{thebibliography}{10}

\bibitem{kreutz2015software}
Kreutz Diego, Ramos Fernando~MV, Verissimo Paulo~Esteves, Rothenberg
  Christian~Esteve, Azodolmolky Siamak, Uhlig Steve. Software-defined
  networking: A comprehensive survey.  {\it Proceedings of the IEEE.
  }2015;103(1);14--76.

\bibitem{mckeown2008openflow}
McKeown Nick, Anderson Tom, Balakrishnan Hari, et al. OpenFlow: enabling
  innovation in campus networks.  {\it ACM SIGCOMM Computer Communication
  Review. }2008;38(2);69--74.

\bibitem{lantz2010network}
Lantz Bob, Heller Brandon, McKeown Nick. A network in a laptop: rapid
  prototyping for software-defined networks.  In: Proceedings of the 9th ACM
  SIGCOMM Workshop on Hot Topics in Networks; 2010; Monterey, California, USA.

\bibitem{wang2013estinet}
Wang Shie-Yuan, Chou Chih-Liang, Yang Chun-Ming. EstiNet openflow network
  simulator and emulator.  {\it IEEE Communications Magazine.
  }2013;51(9);110--117.

\bibitem{son2015cloudsimsdn}
Son Jungmin, Dastjerdi Amir~Vahid, Calheiros Rodrigo~N, Ji~Xiaohui, Yoon Young,
  Buyya Rajkumar. Cloudsimsdn: Modeling and simulation of software-defined
  cloud data centers.  In: Cluster, Cloud and Grid Computing (CCGrid), 15th
  IEEE/ACM International Symposium on; 2015; Shenzhen, China.

\bibitem{vavilapalli2013apache}
Vavilapalli Vinod~Kumar, Murthy Arun~C, Douglas Chris, et al. Apache hadoop
  yarn: Yet another resource negotiator.  In: Proceedings of the 4th annual
  Symposium on Cloud Computing; 2013; Santa Clara, California, USA.

\bibitem{dean2008mapreduce}
Dean Jeffrey, Ghemawat Sanjay. MapReduce: simplified data processing on large
  clusters.  {\it Communications of the ACM. }2008;51(1);107--113.

\bibitem{calheiros2011cloudsim}
Calheiros Rodrigo~N, Ranjan Rajiv, Beloglazov Anton, De~Rose C{\'e}sar~AF,
  Buyya Rajkumar. CloudSim: a toolkit for modeling and simulation of cloud
  computing environments and evaluation of resource provisioning algorithms.
  {\it Software: Practice and Experience. }2011;41(1);23--50.

\bibitem{zeng2017iotsim}
Zeng Xuezhi, Garg Saurabh~Kumar, Strazdins Peter, Jayaraman Prem~Prakash,
  Georgakopoulos Dimitrios, Ranjan Rajiv. IOTSim: A simulator for analysing IoT
  applications.  {\it Journal of Systems Architecture. }2017;72;93--107.

\bibitem{INET}
INET Framework. \url{https://inet.omnetpp.org/}.
\newblock Accessed July 19, 2018.

\bibitem{Floodlight}
Floodlight. \url{http://www.projectfloodlight.org/floodlight/}.
\newblock Accessed July 19, 2018.

\bibitem{garg2011networkcloudsim}
Garg Saurabh~Kumar, Buyya Rajkumar. Networkcloudsim: Modelling parallel
  applications in cloud simulations.  In: Utility and Cloud Computing (UCC),
  Fourth IEEE International Conference on; 2011; Victoria, NSW, Australia.

\bibitem{chen2012workflowsim}
Chen Weiwei, Deelman Ewa. Workflowsim: A toolkit for simulating scientific
  workflows in distributed environments.  In: E-science (e-science), IEEE 8th
  International Conference on; 2012; Chicago, IL, USA.

\bibitem{kliazovich2012greencloud}
Kliazovich Dzmitry, Bouvry Pascal, Khan Samee~Ullah. GreenCloud: a packet-level
  simulator of energy-aware cloud computing data centers.  {\it The Journal of
  Supercomputing. }2012;62(3);1263--1283.

\bibitem{nunez2012icancloud}
N{\'u}{\~n}ez Alberto, V{\'a}zquez-Poletti Jose~L, Caminero Agustin~C,
  Casta{\~n}{\'e} Gabriel~G, Carretero Jesus, Llorente Ignacio~M. iCanCloud: A
  flexible and scalable cloud infrastructure simulator.  {\it Journal of Grid
  Computing. }2012;10(1);185--209.

\bibitem{neves2015mremu}
Neves Marcelo~Veiga, De~Rose Cesar~AF, Katrinis Kostas. MRemu: An
  Emulation-based Framework for Datacenter Network Experimentation using
  Realistic MapReduce Traffic.  In: Modeling, Analysis and Simulation of
  Computer and Telecommunication Systems (MASCOTS), IEEE 23rd International
  Symposium on; 2015; Atlanta, GA, USA.

\bibitem{jung2012mr}
Jung Jongtack, Kim Hwangnam. MR-CloudSim: Designing and implementing MapReduce
  computing model on CloudSim.  In: ICT Convergence (ICTC), International
  Conference on; 2012; Jeju Island, South Korea.

\bibitem{hammoud2010mrsim}
Hammoud Suhel, Li~Maozhen, Liu Yang, Alham Nasullah~Khalid, Liu Zelong. MRSim:
  A discrete event based MapReduce simulator.  In: Fuzzy Systems and Knowledge
  Discovery (FSKD), Seventh International Conference on; 2010; Yantai, China.

\bibitem{wang2009using}
Wang Guanying, Butt Ali~R, Pandey Prashant, Gupta Karan. Using realistic
  simulation for performance analysis of mapreduce setups.  In: Proceedings of
  the 1st ACM workshop on Large-Scale system and application performance; 2009;
  Garching, Germany.

\bibitem{alwasel2017programming}
Alwasel Khaled, Li~Yinhao, Jayaraman Prem~Prakash, Garg Saurabh, Calheiros
  Rodrigo~N, Ranjan Rajiv. Programming sdn-native big data applications:
  Research gap analysis.  {\it IEEE Cloud Computing. }2017;4(5);62--71.

\bibitem{ApacheKafka}
Apache Kafka. \url{http://kafka.apache.org}.
\newblock Accessed July 19, 2018.

\bibitem{HDFS}
Apache Hadoop HDFS. \url{https://hortonworks.com/apache/hdfs}.
\newblock Accessed July 19, 2018.

\bibitem{openvswitch}
Open vSwitch. \url{https://www.openvswitch.org}.
\newblock Accessed July 19, 2018.

\bibitem{dijkstra1959note}
Dijkstra Edsger~W. A note on two problems in connexion with graphs.  {\it
  Numerische mathematik. }1959;1(1);269--271.

\bibitem{guo2009bcube}
Guo Chuanxiong, Lu~Guohan, Li~Dan, et al. BCube: a high performance,
  server-centric network architecture for modular data centers.  {\it ACM
  SIGCOMM Computer Communication Review. }2009;39(4);63--74.

\bibitem{teixeira2013datacenter}
Teixeira Jos{\'e}, Antichi Gianni, Adami Davide, Del~Chiaro Alessio, Giordano
  Stefano, Santos Alexandre. Datacenter in a box: Test your sdn
  cloud-datacenter controller at home.  In: Software Defined Networks (EWSDN),
  Second European Workshop on; 2013; Berlin, Germany.

\bibitem{nardelli2017osmotic}
Nardelli Matteo, Nastic Stefan, Dustdar Schahram, Villari Massimo, Ranjan
  Rajiv. Osmotic flow: Osmotic computing+ iot workflow.  {\it IEEE Cloud
  Computing. }2017;4(2);68--75.

\bibitem{gubbi2013internet}
Gubbi Jayavardhana, Buyya Rajkumar, Marusic Slaven, Palaniswami Marimuthu.
  Internet of Things (IoT): A vision, architectural elements, and future
  directions.  {\it Future generation computer systems. }2013;29(7);1645--1660.

\end{thebibliography}

\end{document}